\documentclass[aps,prb,twocolumn,amsmath,amssymb,superscriptaddress,bibnotes]{revtex4-2}

\usepackage{graphicx}
\usepackage{dcolumn}
\usepackage{bm}
\usepackage{color}
\usepackage{ulem}

\begin{document}

\title{Uniform electronic states and $s$-wave superconductivity in a strongly disordered high-entropy compound (RuRhPdIr)$_{0.6}$Pt$_{0.4}$Sb}

\author{Yufu~Yamada}
\author{Shunsaku~Kitagawa}
\email{kitagawa.shunsaku.8u@kyoto-u.ac.jp}
\author{Taishi~Ihara}
\author{Kenji~Ishida}
\affiliation{Department of Physics, Graduate School of Science, Kyoto University, Kyoto 606-8502, Japan}

\author{Naoto~Uematsu}
\author{Daigorou~Hirai}
\author{Koshi~Takenaka}
\affiliation{Department of Applied Physics, Nagoya University, Nagoya 464–8603, Japan }

\date{\today}

\begin{abstract}
High-entropy compounds, where multiple elements occupy a single crystallographic site in a highly disordered manner, challenge conventional understandings of electronic structure based on periodicity and well-defined band dispersion. 
Here, we report a detailed nuclear magnetic resonance study of the high-entropy superconductor (RuRhPdIr)$_{0.6}$Pt$_{0.4}$Sb, revealing a spatially homogeneous electronic environment in the normal state, in stark contrast to its crystallographically disordered lattice.
The superconducting state exhibits a small but solid Hebel-Slichter coherence peak followed by a significant decrease in the nuclear spin-lattice relaxation rate, providing compelling evidence for fully gapped $s$-wave pairing.
Our findings not only deepen the understanding of superconductivity in highly disordered quantum materials but also open a new pathway for exploring novel superconducting states in entropy-stabilized systems.
\end{abstract}

\maketitle

High-entropy compounds (HECs) are materials in which five or more elements are randomly distributed on a single crystallographic site~\cite{Yeh2004,Zhang2014,Miracle2017,Pickering2016}. 
Unlike conventional materials, where the lowest enthalpy phase is energetically preferred, HECs stabilize alternative crystal structures through high configurational entropy at high temperatures~\cite{Rost2015,Senkov2015,Zhang2016}. 
These structures can be retained as metastable phases at lower temperatures, leading to novel physical and chemical properties.
One of the most remarkable aspects of HECs is the emergence of unprecedented crystal structures that are otherwise inaccessible in conventional materials~\cite{Zhang2014}. 
Furthermore, the complex multi-elemental composition gives rise to the so-called ``cocktail effect'', which can induce unconventional electronic, magnetic, and superconducting (SC) properties~\cite{Hsu2024,Troparevsky2015,Yao2017,Mizuguchi2023}. 
In superconductors, HECs exhibit unique features such as significantly enhanced upper critical fields ($H_{c2}$) and robustness against pressure~\cite{Kozelj2014,Guo2017,Jasiewicz2019,Sun2019,Sobota2022,Pristaifmmode2023,Zeng2024,li2025,sharma2025}. 
However, the microscopic mechanisms underlying these phenomena remain largely unexplored.
A major challenge in understanding the electronic properties of HECs arises from their inherent atomic disorder. 
The random occupancy of multiple elements on a single site makes it difficult to define a periodic unit cell, thereby complicating conventional band structure analyses~\cite{Bansil1999,Stopa2004,Troparevsky2015}. 
As a result, fundamental questions regarding the nature of the electronic states in HECs remain unresolved. 
Despite growing interest in these materials, detailed microscopic investigations, particularly of their electronic and SC properties, are still lacking.

\begin{figure}[!tb]
\centering
\includegraphics[width=\linewidth,clip]{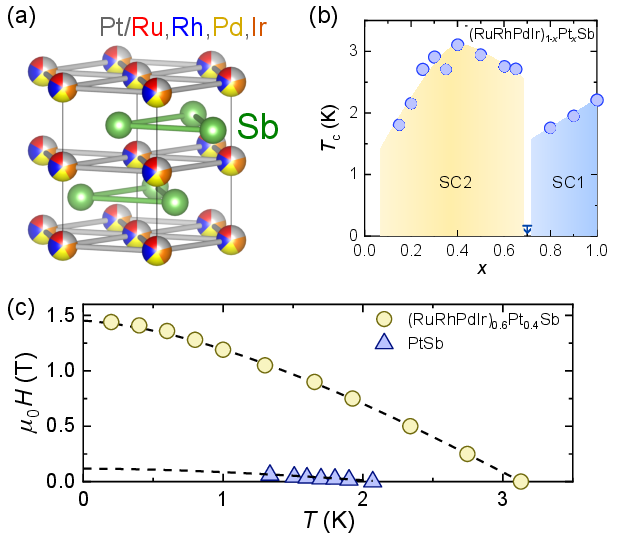}
\caption{
(a) Crystal structure of (RuRhPdIr)$_{1-x}$Pt$_{x}$Sb drawn by VESTA~\cite{K.Momma_JAC_2011}.
Ru, Rh, Pd, Ir and Pt atoms are randomly distributed on the Pt sites.
(b) $x$--$T$ phase diagram of (RuRhPdIr)$_{1-x}$Pt$_{x}$Sb~\cite{Hirai2024}.
Two superconducting phases are separated at $x = 0.7$.
(c) $H$--$T$ phase diagram of PtSb and (RuRhPdIr)$_{0.6}$Pt$_{0.4}$Sb determined by ac susceptibility measurements.
Dashed curves are guides for the eye.
}
\label{Fig.1}
\end{figure}

Recently, some of the authors discovered a new high-entropy superconductor, (RuRhPdIr)$_{1-x}$Pt$_{x}$Sb~\cite{Hirai2023}. 
The parent compound PtSb, crystallizes in a hexagonal NiAs-type structure and exhibits superconductivity with a transition temperature of $T_{\rm c} = 2.1$~K. 
Upon partial substitution of Pt with platinum-group transition metals, (RuRhPdIr)$_{1-x}$Pt$_{x}$Sb retains the NiAs-type structure [Fig.~\ref{Fig.1}(a)] across all compositions, in contrast to related compounds such as RhSb, which adopt a distorted MnP-type structure~\cite{Hirai2024}. 
Interestingly, as shown in Fig.\ref{Fig.1}(b), the SC transition temperature of (RuRhPdIr)$_{1-x}$Pt$_{x}$Sb exhibits a non-monotonic evolution with composition $x$, forming a SC dome with a peak at $x = 0.4$ ($T_{\rm c} = 3.1$~K)~\cite{Hirai2024}. 
However, specific heat measurements suggest that the electronic density of states and Debye temperature do not follow the same trend, implying a possible unconventional SC mechanism driven by high-entropy effects~\cite{Hirai2024}.

In this paper, to gain deeper insight into the electronic and SC properties of (RuRhPdIr)$_{0.6}$Pt$_{0.4}$Sb, we conducted a comprehensive study using nuclear magnetic resonance (NMR) and nuclear quadrupole resonance (NQR) spectroscopy.
These techniques provide powerful probes of both normal and SC states from a microscopic point of view.
Our results reveal that, despite significant atomic disorder, the electronic states remain spatially homogeneous.
Moreover, in the SC state, we observed a small but solid coherence peak, suggesting a full-gap $s$-wave superconductivity in (RuRhPdIr)$_{0.6}$Pt$_{0.4}$Sb.
Our findings provide essential insights into the unconventional SC properties of HECs and contribute to a broader understanding of this emerging class of materials.

Polycrystalline samples of PtSb and (RuRhPdIr)$_{0.6}$Pt$_{0.4}$Sb were prepared using the conventional solid-state reaction~\cite{Hirai2023}.
The X-ray diffraction (XRD) peak widths of (RuRhPdIr)$_{0.6}$Pt$_{0.4}$Sb are nearly identical to those of PtSb~\cite{Hirai2024}.
This indicates that, in contrast to the random occupancy of multiple elements at the Pt site, the distribution of lattice constants is relatively small.
$T_{\rm c}$ in each sample was determined with ac susceptibility measurements.
The $^{121}$Sb (nuclear spin $I = 5/2$, gyromagnetic ratio $^{121}\gamma/2\pi = 10.189$~MHz/T, nuclear electric quadrupole moment $^{121}Q = -0.543 \times 10^{-28}$ m$^2$)-, $^{123}$Sb ($I = 7/2$, $^{123}\gamma/2\pi = 5.5175$~MHz/T, $^{123}Q = -0.692 \times 10^{-28}$ m$^2$)- and $^{195}$Pt ($I = 1/2$, $^{195}\gamma/2\pi = 9.153$~MHz/T)-NMR measurements were performed with a conventional spin-echo technique~\cite{R.K.Harris_2001,N.J.Stone_Q_2016}.
The NMR/NQR spectra as a function of the magnetic field (frequency) were obtained using the Fourier transform of a spin–echo signal observed after a radio-frequency pulse sequence at a fixed frequency (magnetic field).
The magnetic field was calibrated using a $^{63}$Cu ($^{63}\gamma/2\pi = 11.285$~MHz/T)-NMR signal with the Knight shift $K_{\rm Cu} = 0.2385$\% from the NMR coil~\cite{Metallicshifts_1977}.
The Knight shift and full width at half maximum (FWHM) were determined by the peak position and width at the half-maximum value of the NMR spectrum, respectively.
A nuclear spin-lattice relaxation rate $1/T_1$ was evaluated by fitting the relaxation curve of the nuclear magnetization after the saturation to a theoretical function for the nuclear spin $I = 1/2$, which is a single exponential function, and that for $I = 5/2$.

\begin{figure}[!tb]
\centering
\includegraphics[width=\linewidth,clip]{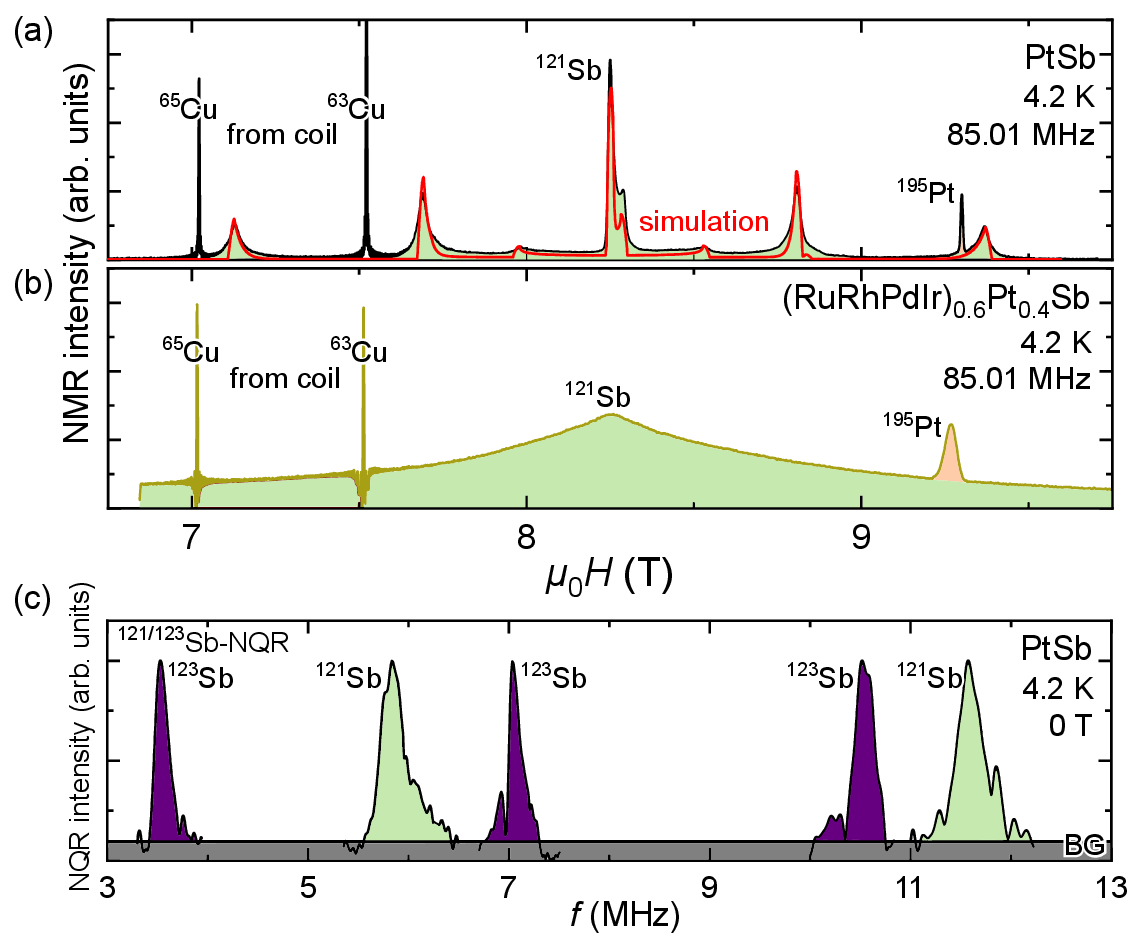}
\caption{
Field-swept NMR spectrum of PtSb (a) and (RuRhPdIr)$_{0.6}$Pt$_{0.4}$Sb (b) measured at 4.2 K and 85.01 MHz.
Numerical simulation assuming a 3:1 ratio between the $c$-axis aligned and randomly oriented components is also plotted (see Supplemental Material~\cite{supplemental}).
(c)$^{121/123}$Sb NQR spectrum in PtSb at 4.2 K.
The spectrum can be reproduced by $^{121}\nu_{zz} = 5.83$ MHz, $^{123}\nu_{zz} = 3.51$ MHz and $\eta = 0$.
}
\label{Fig.2}
\end{figure}

To understand the effects of high-entropy alloying on superconductivity, we first compare the SC phase diagrams of PtSb and (RuRhPdIr)$_{0.6}$Pt$_{0.4}$Sb.
Figure~\ref{Fig.1}(c) shows the magnetic field--temperature ($H$--$T$) phase diagrams determined by ac susceptibility measurements. 
While the upper critical field $H_{\rm c2}$ of PtSb is relatively small ($\sim 0.1$ T), that of (RuRhPdIr)$_{0.6}$Pt$_{0.4}$Sb reaches 1.4~T, approximately 15 times larger, which is consistent with the previous report~\cite{Hirai2024}.
This enhancement is attributed to the reduction of the mean-free path $l$ due to atomic disorder, which effectively shortens the coherence length.
In the orbital limit case, $H_{\rm c2}$ is described as~\cite{Tinkham},
$\mu_0H_{\rm c2} = \frac{\Phi_0}{2\pi \xi^2}$, where $\Phi_0$ is the flux quantum, and $\xi$ is the SC coherence length.
Using this equation, we estimate that the effective coherence length $\xi_{\rm eff}$ is 57 and 15 nm for PtSb and (RuRhPdIr)$_{0.6}$Pt$_{0.4}$Sb, respectively.
In the dirty limit, $\xi_{\rm eff}$ can be described as, $\xi_{\rm eff} = \sqrt{\xi_0l}$, where $\xi_0$ is the SC coherence length in the clean limit.
Assuming $\xi_0$ is the same as $\xi$ in PtSb, $l$ can be estimated to be 3.9 nm, which is one order of magnitude longer than that deduced from transport measurements (0.5 nm)~\cite{Hirai2024}.
This discrepancy is not fully understood and may reflect the complex nature of disorder and scattering processes in high-entropy systems, which warrant further investigation.

\begin{figure*}[!tb]
\centering
\includegraphics[width=15.5cm,clip]{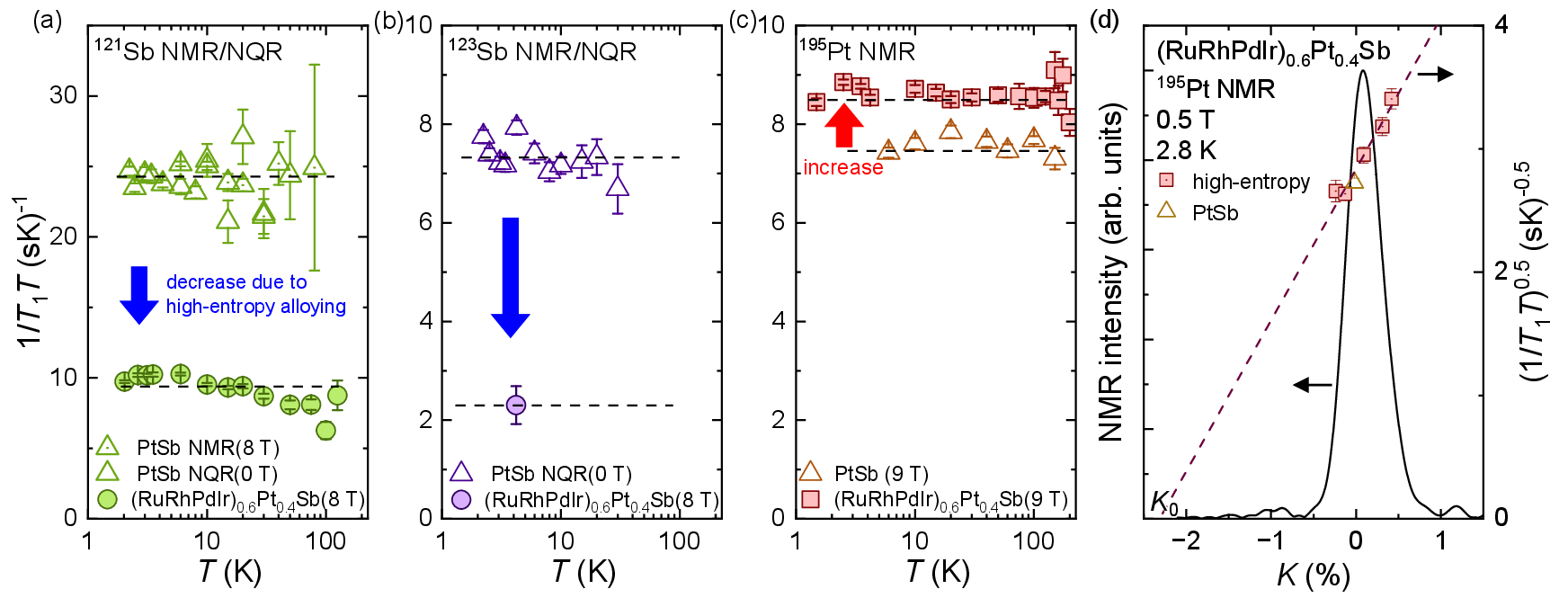}
\caption{
Temperature dependence of PtSb and (RuRhPdIr)$_{0.6}$Pt$_{0.4}$Sb for $^{121}$Sb NMR/NQR (a), $^{123}$Sb NMR/NQR (b), and $^{195}$Pt NMR (c).
As the signal intensity of the $^{123}$Sb NMR in (RuRhPdIr)$_{0.6}$Pt$_{0.4}$Sb is weak, measurements were performed only at a single temperature.
Dashed lines are guides for the eye.
(d) $K$ dependence of square root of $1/T_1T$ together with $^{195}$Pt NMR spectrum at 2.8 K and 0.5 T on (RuRhPdIr)$_{0.6}$Pt$_{0.4}$Sb.
We also plotted the data of PtSb.
The dashed line represents the result of linear fitting.
}
\label{Fig.4}
\end{figure*}

To examine the electronic states in (RuRhPdIr)$_{0.6}$Pt$_{0.4}$Sb, we performed NMR measurements.
Figures~\ref{Fig.2} (a) and (b) show the NMR spectrum of PtSb and (RuRhPdIr)$_{0.6}$Pt$_{0.4}$Sb at 4.2~K, respectively.
The wide-range signal comes from $^{121}$Sb-NMR, while the sharp peak at $\sim$ 9.3 T corresponds to the $^{195}$Pt signal. 
In PtSb, the $^{121}$Sb NMR spectrum consists of five well-defined peaks, indicating a uniform electric and magnetic environment.
However, the NMR spectrum cannot be explained by the NQR parameters ($^{121}\nu_{zz} = 5.83$ MHz and $\eta = 0$) estimated from the NQR spectrum shown in Fig.~\ref{Fig.2}(c).
This discrepancy arises from the sample alignment induced by the magnetic field.
To evaluate the degree of alignment, we performed numerical simulations of the NMR spectrum, as shown in the Supplemental Material~\cite{supplemental}.
As shown in Fig.~\ref{Fig.2} (a), the experimentally observed NMR spectrum is well reproduced by a simulated spectrum (red curves) assuming a 3:1 ratio between the $c$-axis aligned and randomly oriented components.
This result indicates that the magnetic easy axis is parallel to the principal axis of the electric field gradient ($c$-axis).

In contrast to the well-defined $^{121}$Sb NMR spectrum in PtSb, that of (RuRhPdIr)$_{0.6}$Pt$_{0.4}$Sb exhibits a single broad peak, suggesting that the electric field gradient at the Sb nuclei is highly inhomogeneous due to the random occupancy of multiple elements at the Pt site. 
However, the $^{195}$Pt NMR spectrum remains relatively sharp, although slightly broadened compared to PtSb [FWHM at $\sim$ 9.3 T is 7 and 35~mT for PtSb and (RuRhPdIr)$_{0.6}$Pt$_{0.4}$Sb, respectively].
Note that the FWHM scales with the applied magnetic field and becomes sharper at lower fields (FWHM at 0.5~T is 2.5~mT).
A similar or even broader $^{195}$Pt NMR spectrum has been reported in other Pt-containing superconductors~\cite{A.M.Mounce_PRL_2015,Y.Z.Zhou_PRL_2023}, supporting that (RuRhPdIr)$_{1-x}$Pt$_{x}$Sb maintains a macroscopically uniform electronic state.
$^{121}$Sb has a nuclear spin $I \ge 1$ and is affected by electric quadrupole interactions, while Pt has $I=1/2$ and is only affected by magnetic interactions.
Therefore, this observation indicates that despite structural inhomogeneity, the electronic states in (RuRhPdIr)$_{0.6}$Pt$_{0.4}$Sb are spatially uniform across the sample.
The nuclear spin-lattice relaxation rate $1/T_1$ also provides evidence for electronic uniformity, as shown in Fig. S2~\cite{supplemental}. 

\begin{figure}[!tb]
\centering
\includegraphics[width=\linewidth,clip]{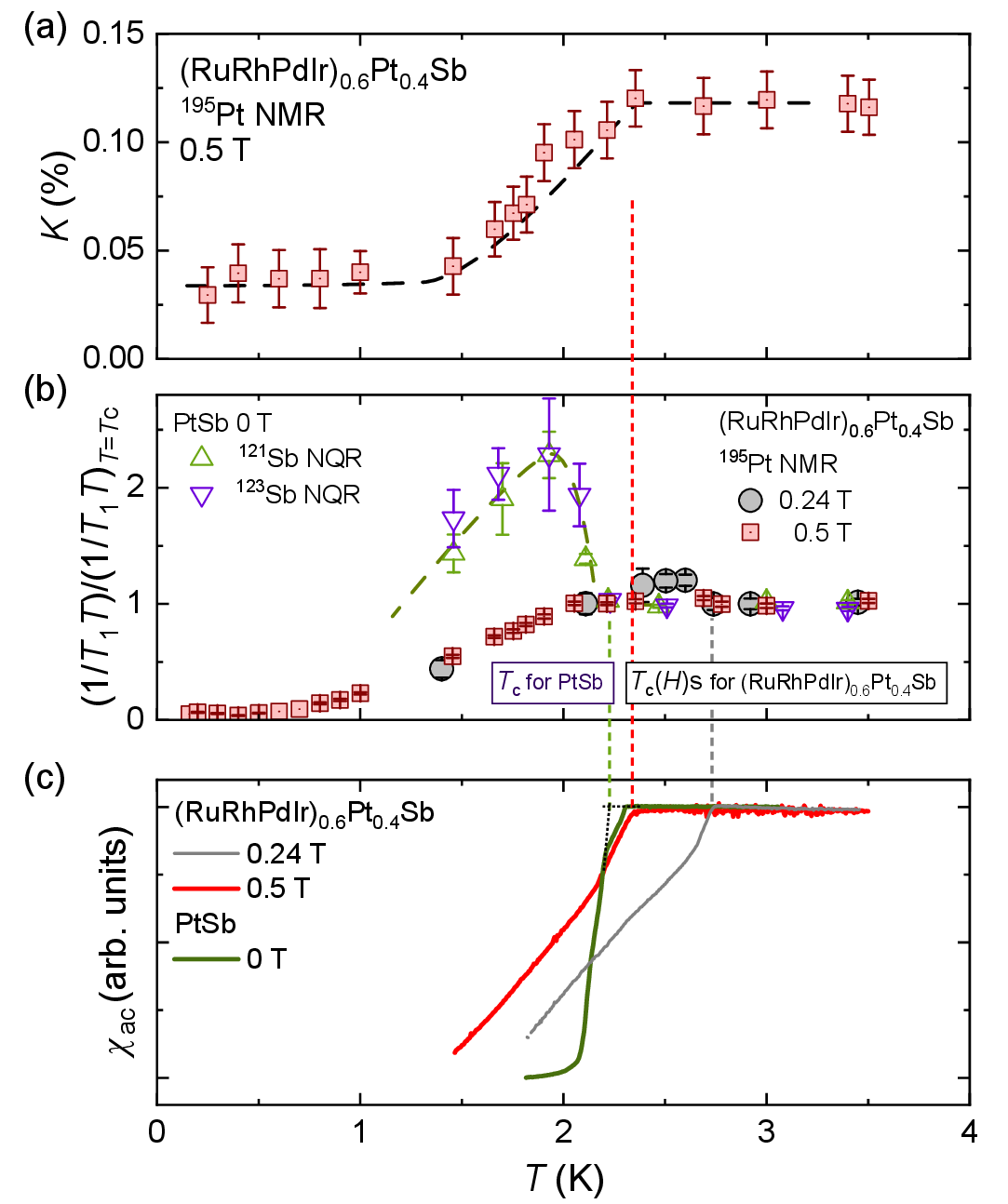}
\caption{
Temperature dependence of Knight shift $K$ at 0.5 T (a), normalized $1/T_1T$ at 0.24 and 0.5 T (b), $\chi_{\rm ac}$ at 0.24 and 0.5 T (c) for (RuRhPdIr)$_{0.6}$Pt$_{0.4}$Sb.
We also plot the data of PtSb for comparison.
The dashed lines indicate $T_{\rm c}$s.
The dashed curves are the guide for the eyes.
}
\label{Fig.5}
\end{figure}

Figures~\ref{Fig.4}(a)--(c) show the temperature dependence of $1/T_1T$ for $^{121/123}$Sb-NMR/NQR and $^{195}$Pt-NMR. 
In both cases, $1/T_1T$ remains constant in the normal state, indicating a Fermi liquid ground state even in highly disordered systems. 
In conventional metals, $1/T_1T$ is proportional to the square of the density of states at the Fermi level $D(E_{\rm F})$.
However, in (RuRhPdIr)$_{0.6}$Pt$_{0.4}$Sb, $1/T_1T$ for Pt increases while that for Sb decreases compared to PtSb.
This suggests that $1/T_1T$ is not simply governed by $D(E_{\rm F})$, but is influenced by additional factors, such as the local distribution of $D(E_{\rm F})$ and quadrupolar interactions unique to high-entropy compounds.
Here, the ratio $^{121}(1/T_1T)/^{123}(1/T_1T)$ is $\sim$ 3.3 for PtSb and $\sim$ 4 for (RuRhPdIr)$_{0.6}$Pt$_{0.4}$Sb, which is close to the value of $(^{121}\gamma/^{123}\gamma)^2 = 3.41$.
This indicates that the relaxation process is predominantly governed by magnetic interactions.

To estimate the spin component of the Knight shift, we examined the correlation between the $^{195}$Pt-NMR Knight shift and the nuclear spin-lattice relaxation rate divided by temperature, $1/T_1T$.
Figure~\ref{Fig.4}(d) displays $\sqrt{1/T_1T}$ as a function of the Knight shift $K = (f - f_0)/f_0$ at 2.8~K, where $f_0$ denotes the reference frequency.
A linear increase of $\sqrt{1/T_1T}$ with increasing $K$ is clearly observed.
In general, the Knight shift can be expressed as
\begin{align}
K = K_{\rm spin} + K_{0},
\label{eq.K}
\end{align}
where $K_{\rm spin}$ denotes the spin contribution, and $K_0$ is the temperature-independent component of the Knight shift.
In conventional metals, $K_{\rm spin}$ is proportional to the spin susceptibility, which in turn scales with the density of states at the Fermi energy $D(E_{\rm F})$.
The observed linear relationship between $\sqrt{1/T_1T}$ and $K$ thus implies that the variation in $1/T_1T$ originates from changes in $D(E_{\rm F})$ at the $^{195}$Pt site.
Remarkably, when the $1/T_1T$ value at the PtSb peak is plotted on the same graph, it falls on the same linear trend, indicating a consistent relationship.
These findings suggest that $D(E_{\rm F})$ in (RuRhPdIr)$_{0.6}$Pt$_{0.4}$Sb is 6\% larger than that in PtSb, even though the residual electronic specific heat coefficient remains unchanged.
This enhancement of $D(E_{\rm F})$, together with a 10\% increase in the Debye frequency and a stronger electron-phonon coupling inferred from the specific heat jump~\cite{Hirai2024}, is likely a key factor contributing to the increase in $T_{\rm c}$ in (RuRhPdIr)$_{0.6}$Pt$_{0.4}$Sb.
From the linear extrapolation of the $\sqrt{1/T_1T}$ vs.\ $K$ plot, the Knight shift corresponding to $\sqrt{1/T_1T} = 0$ yields $K_{0} = -2.3\%$.
The origin of this large negative $K_{0}$ remains unclear.
Typically, $K_0$ is associated with the orbital contribution to the Knight shift $K_{\rm orb}$, which is positive in most metallic systems.
However, negative values of $K_{\rm orb}$ have been reported in Dirac electron systems, where they arise from interband topological effects~\cite{Fukuyama1970,S.Suetsugu_PRB_2021}.
The presence of topological bands has been discussed in other high-entropy superconductors~\cite{Zeng2024}, and future band structure calculations are highly anticipated.
The observed large negative $K_0$ thus indicates an anomalous orbital susceptibility, possibly influenced by topological band structure or strong spin-orbit coupling.

We now focus on the SC state. 
Figure~\ref{Fig.5} (a) shows the temperature dependence of the Knight shift in the SC state of (RuRhPdIr)$_{0.6}$Pt$_{0.4}$Sb.
Below $T_{\rm c}$, the Knight shift decreases by 0.08\%, which is significantly smaller than the estimated spin component $K_{\text{spin}} = K - K_{\text{0}} = 2.4\%$.
This suggests that spin susceptibility remains nearly unchanged in the SC state.

\begin{figure}[!tb]
\centering
\includegraphics[width=7 cm,clip]{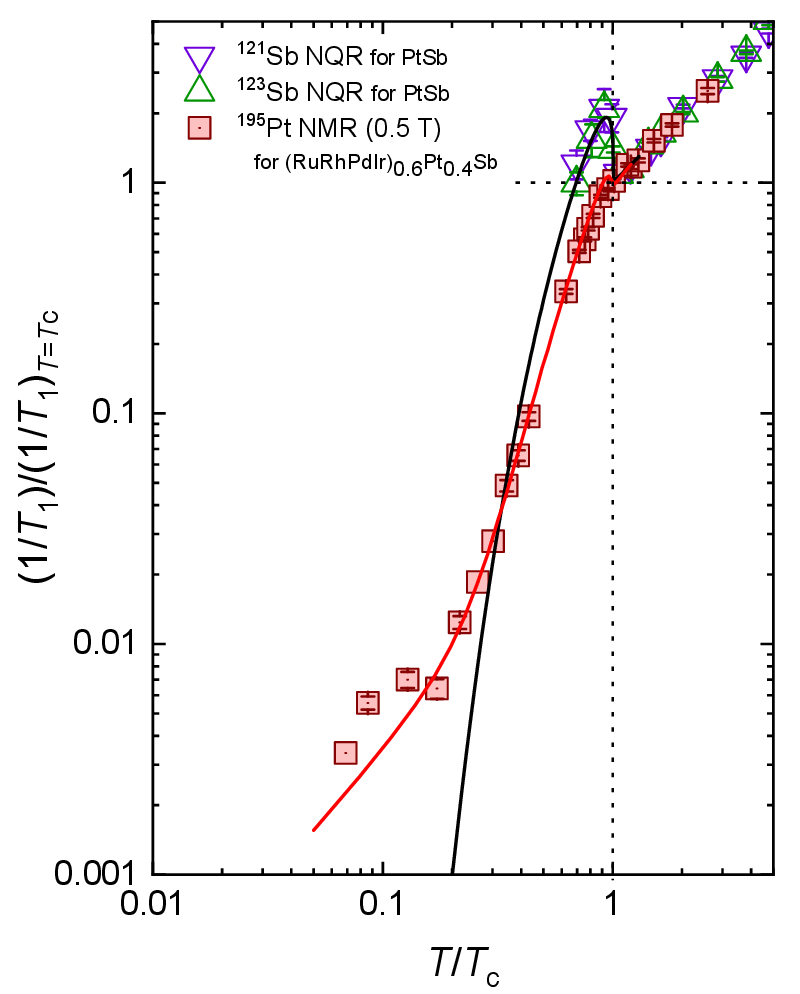}
\caption{
$1/T_1$ normalized by the value at $T_{\rm c}$ plotted against $T/T_{\rm c}$ for PtSb and (RuRhPdIr)$_{0.6}$Pt$_{0.4}$Sb.
The numerical calculation using the BCS theory is shown as the solid curves.
}
\label{Fig.6}
\end{figure}

In term of 1/$T_1T$ in the SC state, we observed a clear coherence peak just below $T_{\rm c}$ in zero-field NQR measurements of PtSb, as shown in Fig.~\ref{Fig.5}(b).
The peak reaches nearly twice the normal-state value, which is a hallmark of conventional $s$-wave superconductivity.
On the other hand, in (RuRhPdIr)$_{0.6}$Pt$_{0.4}$Sb, the coherence peak becomes smaller.
It is noteworthy that the temperature at which $1/T_1T$ starts to decrease is significantly lower than $T_{\rm c}$ determined from the ac susceptibility shown in Fig.~\ref{Fig.5}(c), suggesting the presence of a coherence peak.
The suppression of the coherence peak is likely due to the magnetic field effect~\cite{Y.Nakai_JPSJ_2005,Q.P.Ding_PRB_2016,K.Kinjo_JPSJ_2019}.
Actually, as shown in Fig.~\ref{Fig.5}(b), the value of $1/T_1T$ just below $T_{\rm c}$ increases at lower fields (0.24 T), suggesting that the peak of coherence is partially recovered. 
With further decreasing temperature, $1/T_1T$ drastically decreases and saturates below 0.4~K, indicating the presence of magnetic field-induced residual quasiparticle states~\cite{K.Kinjo_JPSJ_2019}.

Here, we evaluate $1/T_1$ in the full-gap SC state.
We observed a coherence peak just below $T_{\rm c}$.
Conventionally, the presence of a coherence peak is considered evidence of full-gap $s$-wave superconductivity.
We fitted the normalized $1/T_1T$ using the equation deduced from the BCS model ~\cite{L.C.Hebel_PR_1959} (see Supplemental Material~\cite{supplemental}).
Here, $\Delta(0)$ and $\delta$ are the size of the SC gap and a gap smearing factor, respectively.
The results for the full-gap fitting with/without coherence factor are shown as solid curves in Fig.~\ref{Fig.6}.
The $1/T_1$ behavior of PtSb is well described by the conventional BCS parameter with $\Delta(0)/k_{\mathrm{B}}T_{\mathrm{c}} = 1.76$, including the coherence factor.
In contrast, the $1/T_1$ data of (RuRhPdIr)$_{0.6}$Pt$_{0.4}$Sb are almost reproduced using the parameters $\Delta(0)/k_{\mathrm{B}}T_{\mathrm{c}} = 2.1$ and $\delta/\Delta(0) = 0.5$, assuming a 30\% residual density of states and excluding the coherence factor.
The 30\% residual density of states almost corresponds to the field-induced density of states at $H/H_{\rm c2} = 0.36$.
In addition, the external magnetic field suppresses the coherence factor in the $1/T_1$~\cite{Q.P.Ding_PRB_2016}.
The $s$-wave nature of the superconductivity is consistent with the weak pair-breaking effect even in a highly disordered state, as expected from the Anderson theorem~\cite{Anderson1959}.
The extracted SC gap size at $x$ = 0.4 is larger than the weak-coupling BCS value of 1.76 and is consistent with the large specific heat jump $\Delta C/\gamma T_{\rm c} \sim$ 2.3~\cite{Hirai2024}.

Next, we discuss the implications of the reduced Knight shift in the SC state. 
The total Knight shift reduction consists of two components: spin susceptibility suppression $K_{\text{spin}}$ and SC diamagnetic effects $K_{\text{dia}}$. 
Magnetic field dependence of $K_{\text{dia}}$ is described as~\cite{E.Brandt_PRB_2003,M.Manago_JPSJ_2017},
\begin{align}
K_{\text{dia}} = -(1-N)\frac{H_{\text{c2}}}{H} \frac{\ln\left(\frac{H_{\text{c2}}}{H}\right)}{4 \kappa^{2}} \times 100 \ (\%),    
\end{align}
where $N$ (= 1/3) is the demagnetization factor, $\kappa = \lambda/\xi$ is the Ginzburg-Landau parameter, and $\lambda$ is the magnetic penetration depth. 
Given that the SC parameters of (RuRhPdIr)$_{0.6}$Pt$_{0.4}$Sb are not fully established, a precise estimation is challenging.
However, if we assume that the observed 0.08\% reduction is entirely due to SC diamagnetism, we estimate $\kappa \sim 30$, leading to a penetration depth $\lambda \sim 450$~nm.
These values are reasonable for typical type-II superconductors~\cite{Annett,R.Prozorov_SST_2006}.

The minimal spin susceptibility reduction raises the possibility of spin-triplet superconductivity~\cite{H.Tou_PRL_1996,K.Matano_NatPhys_2016,J.Yang_SciAdv_2021,S.Ogawa_JPSJ_2023,G.Nakamine_JPSJ_2019,S.Kitagawa_JPSJ_2024}.
However, given the presence of a small but finite coherence peak and the high-entropy nature of the system, a more plausible explanation is strong spin-orbit scattering.
In systems with large spin-orbit coupling, impurity scattering can mix spin up and down states, preventing spin susceptibility suppression even in spin-singlet superconductors~\cite{K.Miyake_PRB_2000}.
This scenario is consistent with the observed insensitivity of $T_{\rm c}$ to nonmagnetic impurities, a characteristic feature of $s$-wave superconductors.

Regardless of the pairing symmetry, the fact that the spin susceptibility remains nearly unchanged in the SC state suggests that Pauli paramagnetic limiting is ineffective~\cite{T.Hattori_JPSJ_2016,M.Manago_JPSJ_2017,H.Matsumura_JPSJ_2023}.
The robustness of superconductivity against external magnetic fields is a key feature of high-entropy superconductors and warrants further theoretical and experimental investigation.

In this study, we investigated the normal and SC states of the high-entropy superconductor (RuRhPdIr)$_{0.6}$Pt$_{0.4}$Sb using NMR and NQR spectroscopy.
Despite the inherent atomic disorder, we found that the electronic states in (RuRhPdIr)$_{1-x}$Pt$_{x}$Sb remains spatially uniform, as evidenced by the relatively sharp Pt NMR spectrum.
This suggests that, even in a system where multiple elements are randomly distributed over a single crystallographic site, a macroscopically homogeneous electronic state can be realized.
In contrast, the Sb NMR spectrum exhibits significant broadening, reflecting inhomogeneities in the local electric field gradient due to the presence of multiple elements.

The SC properties of (RuRhPdIr)$_{0.6}$Pt$_{0.4}$Sb exhibit several intriguing characteristics distinct from those of conventional superconductors.
The upper critical field $H_{c2}$ is significantly enhanced compared to PtSb, which can be attributed to the reduced coherence length resulting from enhanced nonmagnetic impurity scattering.
Although the observed Hebel-Slichter coherence peak of $1/T_1$ in (RuRhPdIr)$_{1-x}$Pt$_{x}$Sb strongly supports fully gapped $s$ wave superconductivity, the high entropy nature of the system may still offer fertile ground for unconventional phenomena such as inhomogeneous pairing or spin-triplet admixture.
Further studies are required to explore these possibilities.

\section*{acknowledgments}
We acknowledge S. Yonezawa and Z. Hiroi for a fruitful discussion. 
This work was supported by Grants-in-Aid for Scientific Research (KAKENHI Grant No. JP20KK0061, No. JP20H00130, No. JP21K18600, No. JP22H04933, No. JP22H01168, No. JP23H04860, No. JP23H01124, No. JP23K22439 and No. JP23K25821) from the Japan Society for the Promotion of Science, by JST FOREST (Grant No. JPMJFR236M) from the Japan Science and Technology Agency, by research support funding from the Kyoto University Foundation, by ISHIZUE 2024 of Kyoto University Research Development Program, by Murata Science and Education Foundation, and by the JGC-S Scholarship Foundation.
Liquid helium is supplied by the Low Temperature and Materials Sciences Division, Agency for Health, Safety and Environment, Kyoto University.
Y.Y. and S.K. equally contributed to this work.

\end{document}